\documentclass[reprint,aps,prl,english,nolongbibliography,superscriptaddress]{revtex4-2}
\usepackage[colorlinks=true, urlcolor=blue, linkcolor=blue, citecolor=blue, pdftex]{hyperref}
\usepackage{xr} 

\usepackage[dvipsnames]{xcolor}
\usepackage[utf8]{inputenc}
\usepackage[normalem]{ulem}
\usepackage{braket}
\usepackage{graphicx}
\usepackage{caption}
\usepackage{subcaption}
\usepackage{multirow}
\usepackage{tikz}
\usetikzlibrary{decorations.markings, shapes.misc}
\usepackage{cancel}
\usepackage[compat=1.1.0]{tikz-feynman}

\usepackage{amsmath}
\usepackage{amsfonts} 
\usepackage[capitalise]{cleveref}

\newcommand{\D}{\mathcal{D}}

\tikzset{
    magnon/.style={thick, postaction={decorate}, decoration={markings, mark=at position 0.55 with {\arrow{>}}}},
    disorder/.style={thick, dashed},
    vertex/.style={circle, fill=black, inner sep=1.5pt},
    ctvertex/.style={cross out, draw=black, thick, inner sep=3pt},
    vgct/.style={rectangle, fill=white, draw=black, thick, inner sep=2.5pt}
}

\begin{document}

\title{Vanishing spin stiffness in weakly disordered two-dimensional Heisenberg ferromagnets}

\author{Jacopo Niedda}
\email{jniedda@ictp.it}
\affiliation{The Abdus Salam ICTP, Strada Costiera 11, 34151 Trieste, Italy}

\author{Aldo Coraggio}
%\email{acoraggi@sissa.it}
\affiliation{SISSA, via Bonomea 265, 34136, Trieste, Italy}
\affiliation{INFN, Sezione di Trieste, Via Valerio 2, 34127 Trieste, Italy}

\author{Giacomo Bracci Testasecca}
%\email{gbraccit@sissa.it}
\affiliation{SISSA, via Bonomea 265, 34136, Trieste, Italy}
\affiliation{INFN, Sezione di Trieste, Via Valerio 2, 34127 Trieste, Italy}

\author{Antonello Scardicchio}
\email{ascardic@ictp.it}
\affiliation{The Abdus Salam ICTP, Strada Costiera 11, 34151 Trieste, Italy}
\affiliation{INFN, Sezione di Trieste, Via Valerio 2, 34127 Trieste, Italy}

\begin{abstract}
We show that a small fraction of antiferromagnetic bonds qualitatively alters the long-wavelength dynamics of two-dimensional Heisenberg ferromagnets. Although the classical ground state remains magnetized, weak bond frustration generates logarithmically correlated spatial fluctuations of the local spin stiffness, despite the microscopic disorder being short ranged. A replica field theory calculation shows that the effective disorder strength grows under coarse graining, while the spin stiffness decreases, yielding anomalously soft magnons with a scale-dependent dynamical exponent $z>2$. Numerical diagonalization of the semiclassical spin-wave Hamiltonian confirms the anomalous low-energy scaling. The flow is toward an infinite-disorder, zero stiffness regime.
\end{abstract}

\maketitle

\emph{Introduction.}
Long-range order and the nature of collective excitations in two-dimensional magnetic materials have attracted growing interest in recent years in both theoretical~\cite{Yan_2011, Kimchi_2018, Liu2018,Dey2020} and experimental \cite{gong2017,huang2017,bedoyapinto2021,cenker2021,Jenkins2022,bae2022,pal2026} Physics. Two-dimensional quantum spin models are also important test beds for quantum simulators: Rydberg-atom arrays, trapped ions, and superconducting-qubit platforms can now implement frustrated spin Hamiltonians with controlled values for the interactions \cite{ebadi2021quantum,scholl2021quantum,chen2023continuous}. 

For vector spins, while the situation of clean systems at finite temperature $T$ is concisely described by the Mermin-Wagner theorem \cite{Auerbach1994}, which forbids the spontaneous breaking of continuous symmetries at any $T>0$, the properties of the zero temperature phases and their existence in presence of quenched disorder -- including spin-glass, random-singlet, and disorder-induced spin-liquid regimes -- remain strongly model dependent and actively debated, see \emph{e.g.}~\cite{fava2024,Liu2025}. Weak bond disorder, in a globally magnetized sample, is ordinarily expected to modify the spin-stiffness parameter, which sets the quadratic dispersion (and hence the density of states) of the low-lying excitations, the Goldstone modes, without changing the hydrodynamic form of the theory \cite{Halperin_1969,Halperin1977}. In this Letter we show that this expectation is not realized: the spin stiffness vanishes at long wavelengths, even for arbitrarily small bond disorder strength, while the sample remains magnetized. The mechanism is the emergence of logarithmically correlated spatial fluctuations of the \emph{local} spin stiffness out of uncorrelated microscopic disorder, which are exactly marginal in two dimensions.

In a series of previous works, using both modern machine-learning techniques and semiclassical approximation to describe the ground state \cite{viteritti2025,bracci2026}, we investigated the disordered Heisenberg model, in which a fraction $p$ of the exchange bonds is antiferromagnetic. Those studies focused primarily on the spin-glass phase arising at $p\simeq 1/2$; at small $p$, they left unexplained the observation that, while the classical ground states remain spontaneously magnetized on all accessible scales, the smallest spin-wave eigenvalues scale anomalously with the system size, with a dynamical exponent $z>2$ \cite{bracci2026}. 

To identify the microscopic origin of this anomaly, we introduce here a local-twist protocol that probes the spatial fluctuations of the spin stiffness in the classical ground state. We find that the spin stiffness is a random field with long-range, logarithmic spatial correlations, despite the underlying bond disorder being uncorrelated: an emergent scale-free structure that we attribute to the long-ranged 
deformations induced by frustrated bonds in two dimensions \cite{Villain_1977,Parker1988,Vannimenus_1989}, in close analogy with defects in elasticity theory~\cite{giamarchi1995,Baluffi2012,Sutton2020}. Using this numerical input, we write an effective field theory of magnons in a quenched random-stiffness background. Through a replica computation we obtain the one-loop renormalization group (RG) flow, showing that stiffness disorder grows under coarse graining while the average spin stiffness decreases, driving the theory toward a strong-disorder regime. The resulting scale-dependent exponent $z>2$ implies a singular magnon density of states at low energy, an observable signature for spectroscopy of two-dimensional magnets and, in perspective, for quantum simulation platforms.

\emph{The microscopic model.} We consider the two-dimensional disordered Heisenberg model at $T=0$ on the square lattice with nearest neighbor interactions:
\begin{equation}
    H = \sum_{\langle i,j \rangle} J_{ij} \vec{S}_i \cdot \vec{S}_j \, ,
    \label{eq:classicalHam}
\end{equation}
where $\vec S_i$ are $SU(2)$ operators in the spin representation $s$ and the sum runs over nearest-neighbor pairs. In the classical limit, $s\to\infty$, they reduce to vectors on the two-sphere, while for $s=1/2$ they are proportional to Pauli matrices. The exchange couplings $J_{ij}$ are independent random variables drawn from the binary distribution $P(J)=(1-p)\delta(J+1)+p\delta(J-1)$, where $p$ is the antiferromagnetic bond concentration.

\begin{figure}[t]
\includegraphics[width=\columnwidth]{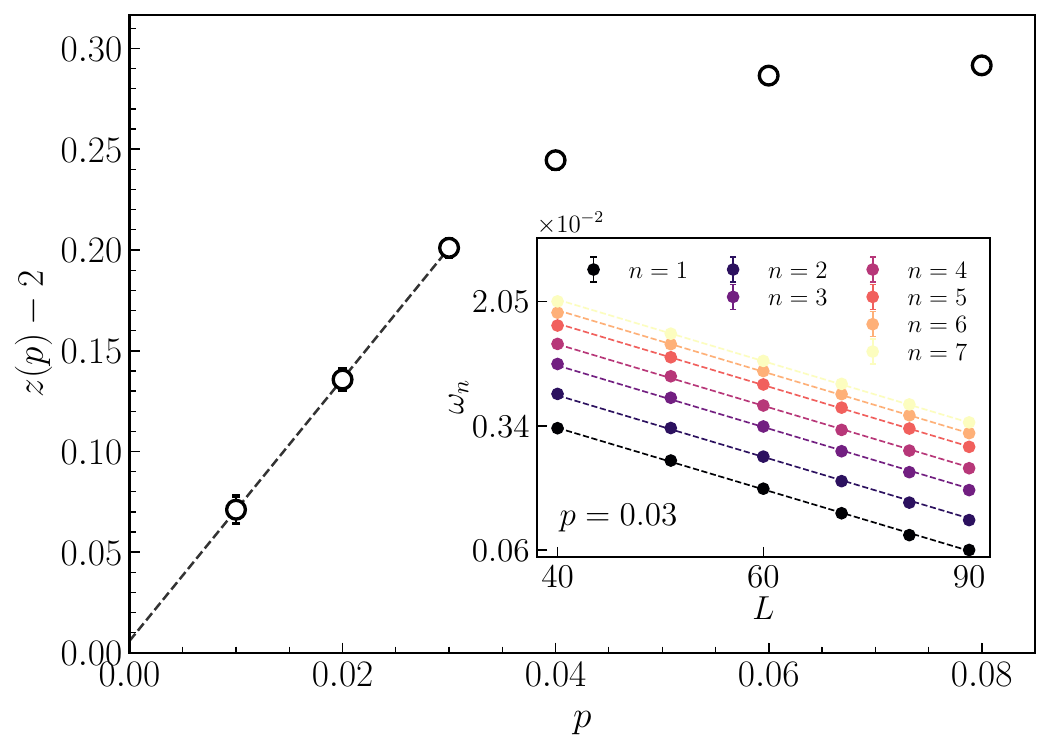}
\caption{Increase of the dynamical exponent $z$ from the clean ferromagnetic values $z=2$ as a function of the antiferromagnetic bond concentration $p$. The linear fit (dashed line) shows the expected behavior for $p \to 0$. (\emph{Inset}) Scaling of $n$-th eigenvalue, averaged over disorder realizations, with respect to the system size $L$, for $p=0.03$.}
\label{fig:spinwaves}
\end{figure}

As we go from $p=0$ to $p=1$ the ground state of the system goes from ferromagnetic (FM) to antiferromagnetic (AFM); at intermediate concentrations the model hosts a quantum spin-glass phase \cite{viteritti2025,bracci2026,oitmaa2001,arrachea2001}. Here we focus on $ p \ll 1/2$, where the classical ground-state configurations are magnetized and weakly canted by the antiferromagnetic bonds. The strict thermodynamic stability of the FM and AFM regions is debated already in the classical limit. A Villain-type argument \cite{Villain_1977} would even suggest that, in the classical model, the FM region is confined to a single point, $p=0$. We do not observe this in our numerical results \cite{bracci2026}. In the worst case scenario (see \emph{End Matter}), the ferromagnetic domains would be of size $L \sim e^{c/p^2}$ with $c\simeq 1$ and, for all practical, numerical, and experimental purposes, they do possess a spontaneous magnetization.

To study the low-energy excitations of~\eqref{eq:classicalHam}, we perform a semiclassical expansion around these classical ferromagnetic canted states. We start with gathering numerical evidence from a $1/s$ Holstein-Primakoff-Bogoljubov calculation.

\emph{Numerical diagonalization of the spin-wave theory.} By following the procedure of~\cite{viteritti2025,bracci2026} and references therein, the quadratic spin-wave (SW) problem can be written in the general Bogoljubov form:
\begin{equation}
\label{eq:HSW}
    H_{SW} = \frac{1}{2} \sum_{ij} \left( A_{ij} a_i^\dagger a_j + B_{ij} a_i^\dagger a_j^\dagger \right) +  \mathrm{H.c.} \; , 
\end{equation}
where $[a_i,a^\dagger_j]=\delta_{ij}$. The matrix elements of $A$ and $B$ are defined in terms of the random couplings $J_{ij}$ and the local rotation matrices aligning each spin of the classical configuration to the same axis (see~\cite{bracci2026} for details). To find the classical energy minima we adopt an \emph{over-relaxation} gradient descent algorithm~\cite{Baity2015inherent,Baity2015soft,bracci2026}.

The low-energy sector of the SW spectrum can be studied by numerically implementing a Bogoljubov transformation to diagonalize $H_{SW}$~\cite{bracci2026,Bogo_1947,Colpa_1978,blaizot_ripka_1986}. For $p \ll 1/2$, where the classical reference states are spontaneously magnetized, the lowest-energy modes are completely delocalized~\cite{bracci2026}. It is therefore sound to assume the existence of a dispersion relation $\omega \sim k^z$ for small wavenumber $k$, where $z$ is the dynamical exponent. A standard computation relates the dispersion relation to the density of states: $\varrho(\omega) \sim \omega^{d/z-1}$, in generic spatial dimension $d$. For the clean FM, $A$ is the discrete Laplacian on the lattice: $H_{SW}^{p=0}=-\sum_{<ij>}a^\dagger_i a_j$, and $B_{ij}=0$. The dispersion relation is $\omega \sim k^2$, and therefore $z=2$. Correspondingly, the density of states is $\varrho(\omega)\sim\omega^0$. 

After the introduction of small bond disorder, we observe, instead, an abundance of low-energy modes corresponding to a softer ($z>2$) dispersion relation, and leading to a singular density of states at zero energy. To determine the dynamical exponent, we study the finite-size scaling of the lowest positive SW frequencies. In Fig.~\ref{fig:spinwaves}, for each mode $n\leq 10$, we fit $\omega_n \sim L^{-z}$, averaged over disorder. The resulting exponents are mode independent, supporting a common dynamical exponent $z$, and are systematically larger than the clean value $z=2$. Moreover, $z$ increases with the concentration $p$ of antiferromagnetic bonds.

%independently of the binning involved in the density of states

\begin{figure*}[t]
\centering
\begin{subfigure}{.49\textwidth}
  \centering
\includegraphics[width=0.7\linewidth]{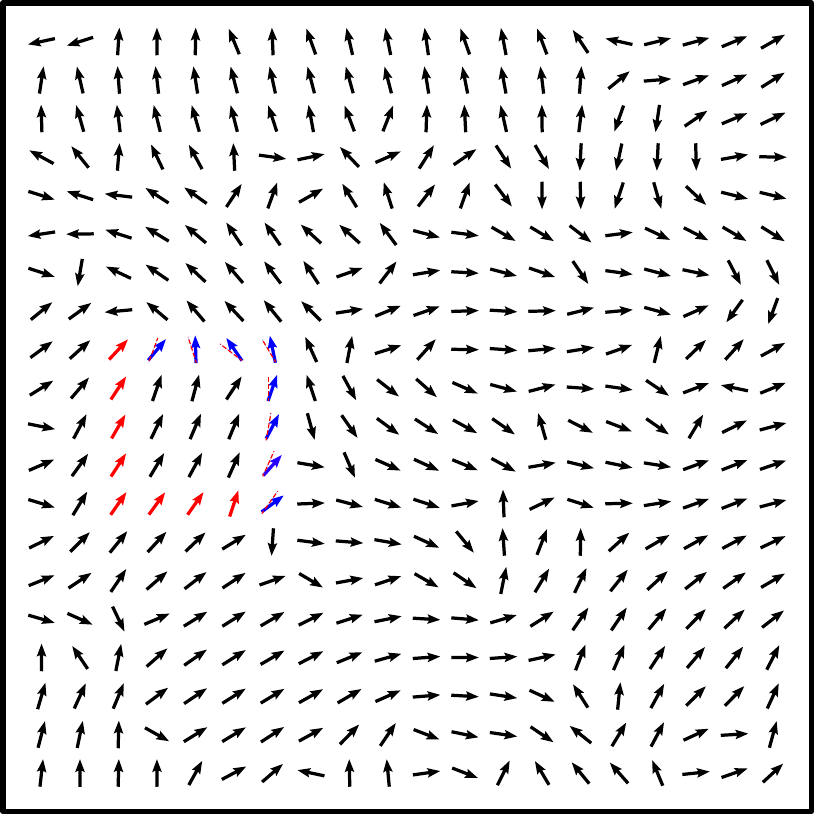}
\end{subfigure}
\begin{subfigure}{.49\textwidth}
  \centering
\includegraphics[width=0.98\textwidth]{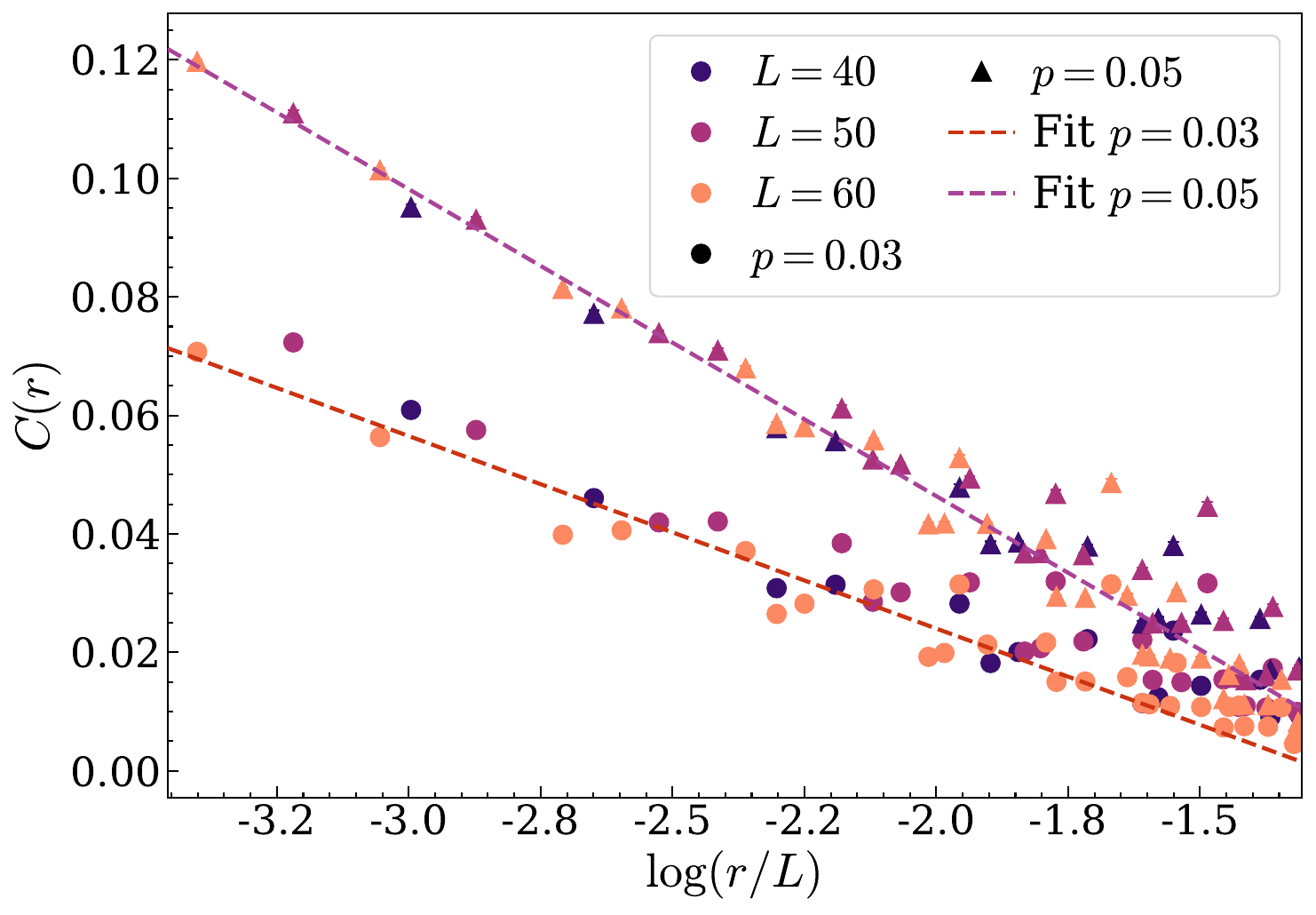}
\end{subfigure}
\caption{\textbf{Left panel:} A square patch on a classical ground state spin-configuration, used to measure the \emph{local} spin stiffness. The spin vectors are three-dimensional, even though they are drawn as planar arrows for clarity. The boundary red spins are kept fixed, while the boundary blue spins are  rotated by a uniform angle $\theta=\pi/10$. Then, all spins on the boundary of the patch are held fixed, while the remaining spins in the system are relaxed to minimize the energy. The value of the stiffness of the clean FM in this protocol is $\rho_s^{(0)} \simeq 3.5$ is an arbitrary scale. \textbf{Right panel}: Spin stiffness correlations as a function of the distance. From the linear fit, we find $\Delta \simeq 0.2$ for $p=0.03$ (for which $\rho_s=2.4$) and $\Delta \sim 0.3$ for $p=0.05$ (for which $\rho_s=2.1$).}
\label{fig:stiff_corr}
\end{figure*}

\emph{Spin-stiffness spatial fluctuations.} In order to make analytic progress, we should take two limits: first the long-wavelength limit, for the low-energy modes of (\ref{eq:HSW}), and second the small $p$ approximation, in which the distance between the microscopic defects is $l\sim 1/p\gg 1$. The matrices $A,B$ are correlated random matrices, which can be computed only after the classical minimum has been found \cite{bracci2026}. In the small $p$ limit, and long wavelength limit $1\ll 1/p\ll L$ one can reason phenomenologically to derive an effective field theory (see \emph{End Matter}) which does not depend explicitly on the lattice constant.

At long wavelengths, we have a complex scalar magnon field $\phi = \phi(\vec x,t)$, with action:
\begin{equation}
\label{eq:ActionM}
    S[\phi] = s \int d^2x dt \left( i\phi^* \partial_t \phi - \rho_s(\vec x) |\nabla \phi|^2 \right) \; ,
\end{equation}
where $\rho_s(\vec x)$ is the local spin-stiffness, and $s \sim 1/\hbar$ is the spin representation. In the weak-disorder regime, the leading effect of the antiferromagnetic bonds is to promote the stiffness constant to a quenched, slowly-varying random field $\rho_s\to\rho_s(\vec x)$. This field can be thought of as the cost of twisting the spins away from their equilibrium position, in a region of space around the point $\vec x$ and must be computed for any classical disorder realization. In order to gain insight into its nature, we have developed an \emph{ad hoc} numerical technique~\footnote{The {\it global} spin stiffness, also referred to as the helicity modulus~\cite{Shastry_1990,Bonca1994}, is usually measured from the free-energy cost associated with an imposed twist of boundary conditions. For a small twist angle $\theta$, we write $F(\theta)-F(0)=\rho_s\theta^2+O(\theta^4)$. At $T=0$, the free energy coincides with the energy $F(\theta)=E(\theta)$.}. 

After having found the classical ground state, we select a square patch and impose a twist of magnitude $\theta$ on the spins along its upper and right boundaries, while those along the lower and left boundaries are kept fixed at their original orientations, as illustrated in Fig.~\ref{fig:stiff_corr}. Then, the energy is minimized once again keeping the patch boundaries fixed. The local stiffness associated with a given patch centered in $\vec x$ is then measured as $\rho_s(\vec x) = \Delta E(\vec x) / \theta^2$, where $ \Delta E(\vec x)$ is the energy increase due to the imposed twist. We used $(L/2)^2$ overlapping $10\times10$ patches in such a way to densely sampling the whole lattice. 

We then measure the spatial correlations of $\rho_s(\vec x)$ separating an average and a fluctuating part $\rho_s(\vec x)=\rho_s+\delta\rho(\vec x)$, with 
$\delta \rho (\vec x)$ zero averaged and correlated over disorder realizations. In the right panel of Fig.~\ref{fig:stiff_corr} we show that  the correlation at fixed distance $r$ is well fitted to decay logarithmically with the distance:
\begin{equation}
    C(\vec x, \vec x') \simeq -\frac{\Delta}{2\pi} \ln \left( \mu  |\vec x - \vec x'| \right) \; , \quad \mu \sim L^{-1} \; ,
    \label{eq:logcorr}
\end{equation}
with an amplitude $\Delta$ increasing with $p$. This measure reveals that spatially uncorrelated bond disorder generates long-range, logarithmically correlated fluctuations of the spin stiffness. As we show below, this is precisely the marginal form of stiffness disorder capable of modifying the infrared magnon dynamics in two dimensions.

\emph{Effective field theory.}  The spin stiffness can be divided into a microscopic average, bare, stiffness $\rho_s$ and its fluctuations $\delta\rho$, discussed in the previous section. Accordingly, the action \eqref{eq:ActionM} can be split into a clean, bare, magnon action \cite{Auerbach1994,Halperin_1969} and an interaction term between magnons and a static random field $\delta \rho(\vec x)$:
\begin{eqnarray}
\label{eq:action}
    S[\phi,\delta\rho] &=& s \int d^2x dt \left( i\phi^* \partial_t \phi - \rho_s |\nabla \phi|^2 \right) \nonumber\\
    &-& s\int d^2x dt~\delta \rho(\vec x) |\nabla \phi|^2 \; .
\end{eqnarray}
We take the quenched random stiffness field to be Gaussian distributed with covariance matrix given in~\eqref{eq:logcorr}: $C(\vec x, \vec x') = \mathbb{E}[\delta\rho(\vec x)\delta\rho(\vec x')]$, where \begin{equation}
    \mathbb{E}[(\cdot)]=\frac{1}{\mathcal{N}}\int \D\delta\rho~e^{-\frac{1}{2}\int d^2x\ d^2x' \delta\rho~C^{-1}~\delta \rho}(\cdot)  \; ,
\end{equation}
with the normalization $\mathcal{N}$ fixed by $\mathbb{E}[1]=1$. The magnon propagator is
\begin{equation}
    G(\vec x,t| \vec x',t')=\langle\phi(\vec x,t)\phi^*(\vec x',t')\rangle \; ,
\end{equation}
where $\langle (\cdot) \rangle$ denotes the average over the measure $\mathcal D \phi$ at fixed $\delta\rho$. The density of states is related to the averaged propagator as:
\begin{equation}
\varrho(\omega)=\mathbb{E}\left[\frac{1}{\pi}\Re\int\frac{d^2k}{(2\pi)^2}G(k,\omega)\right] \; ,
\end{equation}
and its scaling properties can be read by the pole of $G$ at the appropriate wave-number $k$.

\begin{figure}[t]
\centering
\begin{subfigure}{\linewidth}
\centering
\begin{tikzpicture}
    \begin{feynman}
    \vertex (i1) at (0,0);
    \vertex (a) at (2,0);
    \vertex (b) at (5,0);
    \vertex (i2) at (7,0);
    \diagram*{
    (i1) -- [fermion, edge label= \(\vec p \)] (a) -- [fermion, edge label= \( \vec p - \vec q \)] (b) -- [fermion, edge label= \( \vec p  \)] (i2),
    (a) -- [scalar, half left, edge label= \( \vec q  \)] (b),
  };
\end{feynman}
\end{tikzpicture}
\end{subfigure}

\vspace{1.5em} % Adds vertical space between the two diagrams

\begin{subfigure}{\linewidth}
\centering
   \begin{tikzpicture}
   \begin{feynman}
   \vertex (i1) at (0,0);
   \vertex (i2) at (4,0);
   \vertex (a) at (1,1);
   \vertex (b) at (3,1);
   \vertex (c) at (2,2);
   \vertex (i3) at (2,3);
   \diagram*{
    (i1) -- [fermion, edge label= \(\vec p \)] (a) -- [fermion, edge label= \( \vec p - \vec q \)] (c) -- [fermion, edge label= \( \vec p' - \vec q \)] (b) -- [fermion, edge label= \( \vec p'  \)] (i2),
    (a) -- [scalar, edge label= \( \vec q  \)] (b),
    (c) -- [scalar, edge label= \( \vec p -\vec p'  \)] (i3),
  };
\end{feynman}
\end{tikzpicture}
\end{subfigure}
\caption{The two relevant diagrams for the one-loop computation: self energy (top) and vertex/disorder renormalization (bottom). Continuous line represents the bare $\phi$ propagator, while the dashed line is the static disordered $\sigma$ propagator.}
\label{fig:RGdiag}
\end{figure}
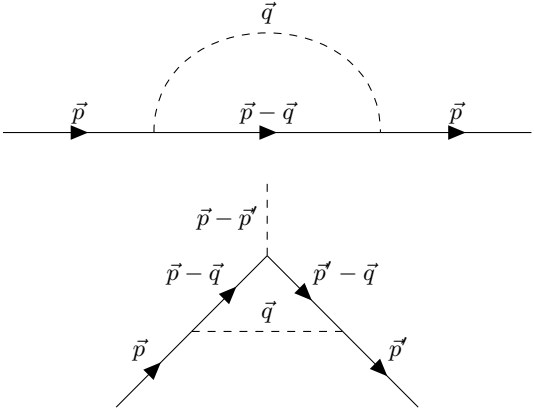

\begin{figure}[t]
\includegraphics[width=0.8\linewidth]{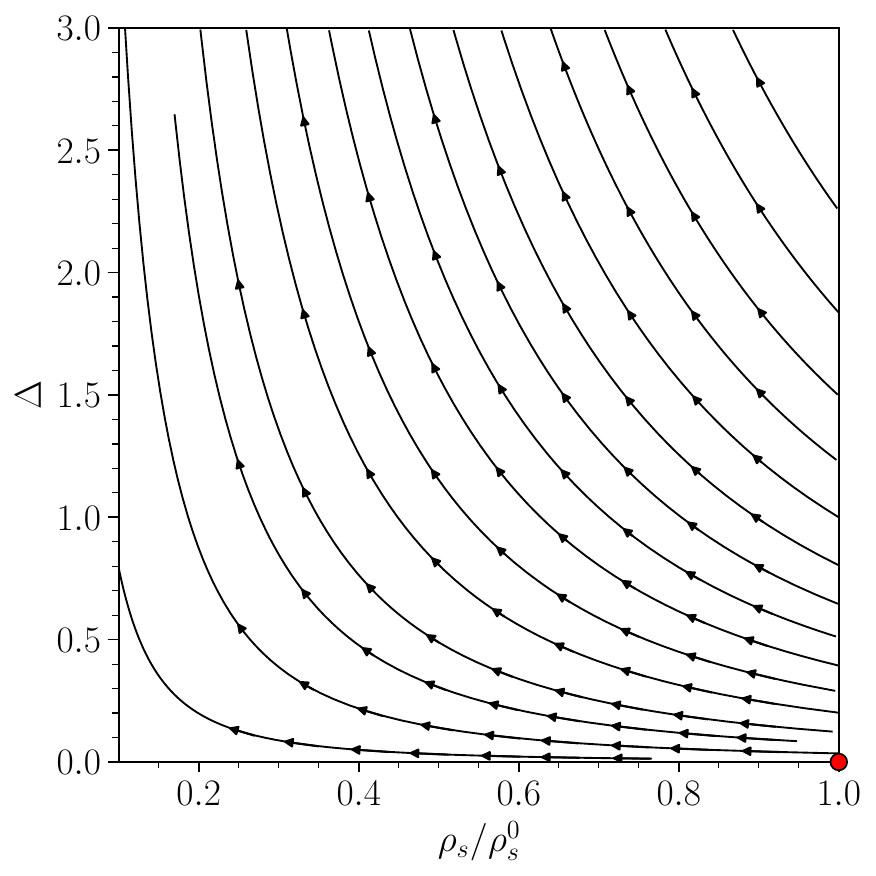}
\caption{One-loop RG flux lines. One can recognize the RG orbits where $\rho_s^2\Delta$ is a constant. The red dot is the \emph{repulsive} FM fixed-point. The system runs to an infinite-disorder regime with $\rho_s=0, \; \Delta\to\infty$.}
\label{fig:one-loop_RGflow}
\end{figure}

To compute disorder-averaged correlation functions, we use the replica trick \cite{Mezard87,Efetov96}. For the propagator, one has:
\begin{equation}
\label{eq:aveprop}
    \mathbb{E}[G(k,\omega)]=\mathbb{E}\left[\langle\phi_1(\vec k,\omega)\phi^*_1(\vec k,\omega) \rangle\right]_{n\to 0} \; ,
\end{equation}
where $n$ is the number of replicas, eventually sent to zero. In the replicated theory, the average over the quenched stiffness disorder generates a nonlocal \emph{quartic} interaction between replicas $\phi_a$, mediated by the static correlator $C(\vec x, \vec x')$. Equivalently, we can rewrite the two averages in \eqref{eq:aveprop} (over quantum fluctuations and over disorder) to be almost on the same footing by defining:
\begin{equation}
\mathbb{E[\langle(\cdot)\rangle]} =\int\mathcal{D}\phi\mathcal{D}\sigma~e^{iS_n[\phi,\sigma]} (\cdot) \; ,
\end{equation}
in terms of a replicated action $S_n[\phi,\sigma] = \sum_{a=1}^n S[\phi_a,\sigma]$ where we have introduced, a normalized, time independent, random field $\sigma(\vec x)$ and a coupling strength $g$ such that $\delta\rho (\vec x)=g\sigma(\vec x)$. We assume the measure $\D \sigma$ to include $\mathbb{E}[(\cdot)]$, so that, to reproduce the numerical observation \eqref{eq:logcorr}, the bare $\sigma$ correlator is: 
\begin{equation}
    \langle\sigma(\vec x)\sigma(\vec x')\rangle_0 = \int \frac{d^2q}{(2\pi)^2}\frac{e^{i\vec{q}\cdot(\vec{x}-\vec{x}')}}{q^2+\mu^2} \; ,
\end{equation}
where $\mu \sim L^{-1}$ acts as an infrared regulator. This expression reduces to \eqref{eq:logcorr} for $|\vec x - \vec x'|\ll 1/\mu\sim L$, and, therefore, $\Delta=g^2$ connects the effective field theory to our numerical evidence on the microscopic model.

It is worth noting that only a logarithmically correlated disorder may affect the infrared theory. Indeed, the bare scaling dimensions in momentum units are: 
$   [x]=-1, \quad [t]=-2, \quad [\phi]=1, \quad [\rho_s]=0.$
A short-range disorder correlator $C(\vec{x},\vec{x}') = \Delta \delta^{(2)}(\vec{x}-\vec{x}')$, would give $[\Delta] = -2$ and $[g]=-1$, meaning that short-range stiffness disorder is an irrelevant coupling~\footnote{More generally, power-law correlations $C(r)\sim \Delta r^{-\alpha}$ would require $[\Delta]=-\alpha$ and are again irrelevant for $\alpha>0$.}. The logarithmically correlated stiffness disorder numerically observed in Fig.~\ref{fig:stiff_corr} is instead marginal at bare level, \emph{i.e.} $[\Delta] = [g] = 0$, and its effect on the infrared theory must be determined beyond power counting. 

\emph{One-loop RG flow.} We have solved the interacting field theory perturbatively with a one-loop RG computation. The relevant diagrams for the self-energy and the vertex function evaluation are the ones drawn in Fig.~\ref{fig:RGdiag}. The details of the computation are sketched in the \emph{End Matter}. By extracting the UV divergent contributions from the diagrams, we find the following expression of the RG-flow equations, written in terms of $\rho_s$ and $\Delta=g^2$:
\begin{subequations}
    \begin{align}
        \frac{d \rho_s}{d\ell} &= - \frac{\Delta}{4 \pi \rho_s} \\
        \frac{d \Delta}{d\ell} &= \frac{\Delta^2}{2 \pi \rho_s^2} \; ,
    \end{align}
\end{subequations}
where $\ell = \ln (L/L_0)$ and $L_0$ is the microscopic starting point (\emph{i.e.}~the size of the plaquettes used to compute the stiffness). While the stiffness decreases, the disorder increases {\it indefinitely}. It is not difficult to see that the combination $\rho_s^2\Delta$ is a constant of the RG flow and this reduces the flow to a \emph{one-parameter} scaling. The RG orbits are shown in Fig.~\ref{fig:one-loop_RGflow}. 

To identify the running coupling and in order to compare with our numerical results, we look at the anomalous dimension of the stiffness, whose RG equation is:
\begin{equation}
    \frac{d\ln\rho_s}{d \ell} =
    -\frac{\Delta( \ell)}{4\pi \rho_s(\ell)^2}
    \equiv -\lambda(\ell) \; .
\end{equation}
It is not difficult to see that $\lambda$ satisfies the autonomous equation $d\lambda/d\ell=4\lambda^2$. The solution $\lambda=\lambda_0/(1-4\lambda_0\ell)$ has a Landau pole at $\ell=1/4\lambda_0$. This non-perturbatively large lengthscale $L^*=L_0\exp(1/4\lambda_0)$, must be considered as a limit of applicability of the one-loop calculations to the microscopic model.

Since the one-loop self-energy does not renormalize the $\phi^*\partial_t\phi$ term (\emph{i.e.}~$Z_\phi=1$ at one loop), the dispersion keeps the form:
\begin{equation}
\label{eq:disp_rel}
    \omega(k) = \rho_s(k) k^2\to k^{2+\lambda(\ell)} \; ,
\end{equation}
where the running coupling $\rho_s$ is computed at the inverse momentum scale, $\ell \simeq - \ln k$, of the modes we are looking at. Eq.~\eqref{eq:disp_rel} yields our main result: a relation between the scale-dependent dynamical exponent $z$ and the dimensionless, running coupling $\lambda$:
\begin{equation}
    z(\ell) = 2 + \lambda(\ell) > 2 \; .
\end{equation}

The identification of the running coupling $\lambda$ with the measured $z-2$ would seem to allow more than a qualitative comparison with the one-loop RG. However, this is premature. For $p=0.03$, the microscopic ($10\times 10$ plaquette) value of $\lambda_0=z(0)-2 \simeq 0.0028$ starts very small and does not significantly grow for $L \leq 90$ since $L\ll L^*\sim 10^{38}$. So, while the RG explains $z>2$ at one loop, a quantitative comparison requires probably going beyond one loop, or identifying other sources of correlated disorder. It is also worth noticing that the measured values of $z-2$ for $L \leq 90$ plateau at around $0.3$ upon increasing $p$ (see Fig.~\ref{fig:spinwaves}). 

\emph{Discussion.}
A finite order parameter coexisting with a vanishing stiffness has analogues in other systems, the amplitude of the order parameter and the associated rigidity being logically independent quantities. In strongly disordered superconducting films, the pairing amplitude survives locally while the superfluid stiffness is suppressed and drives the superconductor-insulator transition \cite{emery1995,ghosal2001}; in two-dimensional melting, unbinding dislocations drive the renormalized shear modulus to zero while local crystalline order persists \cite{halperin1978,nelson1979,young1979}; in pinned flux lattices, quasi-long-range order coexists with anomalous, scale-dependent elasticity \cite{Cardy82, giamarchi1995}. In the superconducting case, however, the loss of rigidity requires strong disorder; here the stiffness flows to zero for arbitrarily weak disorder, as a consequence of the double marginality of two dimensions: the stiffness is dimensionless, and the emergent logarithmically correlated disorder is exactly marginal. The same mechanism should operate in any two-dimensional system with a continuous order parameter and dilute local frustration.

The scale-dependent stiffness also makes contact with the droplet description of spin glasses \cite{Fisher1986,Fisher1987,Fisher_Huse1988}, in which the energy of a system-spanning excitation scales as $\Delta E(L)\sim L^{\theta_{\mathrm{st}}}$. Since the cost of a twist on scale $L$ behaves as $\Delta E(L)\sim\rho_s(L)L^{d-2}$, in $d=2$ the running logarithmic slope gives $\theta_{\mathrm{st}} = -\lambda(\ell) = 2 - z(\ell) < 0$: the same loss of rigidity that, in droplet theory, signals a zero-temperature critical regime, and that has been measured in two-dimensional Ising, XY, and Heisenberg spin glasses \cite{Hartmann2001,Hartmann2008,Kawamura_2003,weigel2006ground}. The correspondence must be qualified: our $\theta_{\mathrm{st}}$ is a linear-response exponent, obtained from the harmonic curvature around a single classical minimum, whereas droplets are $O(1)$ rearrangements between competing minima. The two notions coincide for $L\ll L^*$, where the response to a twist is a smooth canting; the flow toward strong disorder signals the scale at which genuine droplets should take over.

Our result should also remind the educated reader of that of S.~Chakravarty \cite{Chakravarty_1991}, who, motivated by the analogy with the scale-free conductance fluctuations found by Altshuler, Kravtsov, and Lerner \cite{altshuler1986statistics}, considered a clean two-dimensional Heisenberg ferromagnet at $T>0$, on scales smaller than the correlation length. By integrating out the high-energy magnons, Chakravarty found a reduction of the global spin stiffness and a growth of its fluctuations. The reduction found here is similar, but instead of flowing to the paramagnet (the fixed point of the clean ferromagnet at $T>0$), our model flows to an \emph{infinite disorder fixed point} with zero stiffness, reminiscent of a weakly ordered Griffiths phase \cite{vojta2005quantum,mohan2010infinite,vojta2006rare,Laumann2008}.

It is worth mentioning that our results, being confined to the quadratic magnon Hamiltonian, are valid both for the quantum and the {\it classical} Heisenberg model. In fact, while the eigenstates are different, the spectrum of the Bogoljubov Hamiltonian is the same of the dynamical matrix of the classical model. Since the $s=1/2$ case will contain $1/s$ corrections that correspond to magnon interactions, we know from studies of many-body localization \cite{basko2006metal,Sierant_2025} that interactions might destroy localization and restore hydrodynamics. Our work poses the basis for the study of the interacting, small-$s$ model as well.

Although we started (for historical reasons) with the more difficult case of the Heisenberg model, a similar analysis can be performed for the XY model, whose classical configurations have been studied by several authors in the past \cite{cieplak1985scaling,weigel2006ground}. The role of disorder frustration in the XY model has been studied since the seminal work of Villain \cite{Villain_1977}, also by applying renormalization group ideas~\cite{Jose1979,Jose1981}. By building on these results, we plan to extend our analysis to this case in a future work.

\emph{Acknowledgments.} We thank R.\ Moessner, A.\ Pandey, and M.\ Weigel for discussions and collaborations on related topics, and G.\ Villadoro for discussions on the extension of our one-loop results to two and higher loops, although, unfortunately, we could not complete the two-loop calculation in time for publishing it in this paper.

\clearpage

\section*{End Matter}

\subsection{Dipolar deformations of the classical ferromagnetic ground state}
Let us consider the classical Hamiltonian of the ferromagnetic Heisenberg model \eqref{eq:classicalHam}, with $J_{ij}=-1$. The addition of a single, isolated antiferromagnetic bond is borderline for the classical ground state since it does not change the minimum-energy configuration, unless $J>1$ \cite{Parker1988,Vannimenus_1989}. However two $J=1$ bonds at $O(1)$ distance from each other can {\it cant} the ground state, without reducing the overall magnetization too much. So the number of defects that
can actually create a deformation of the magnetization is only $p^2 L^2$ out of the total $p L^2$.

Following Villain \cite{Villain_1977}, we consider the case of dipolar deformations. In the continuum limit, when the defects are well separated, we can write a non-linear sigma model \cite{Auerbach1994} for the local magnetization field $\vec m(\vec r)$:
\begin{equation}
    H = J \sum_{ij} (\vec{m}_i - \vec{m}_j)^2 \sim \int d^2r\sum_{\alpha = \{ x,y,z \}} \left( \nabla m^\alpha \right)^2 \; .
\end{equation}
We choose the reference frame such that in absence of defects $m^z_i=1$, and so we assume that a few defects produce \emph{small} fluctuations in the orthogonal plane: $|m^{x,y}_i|\ll 1$. In the ground state, the magnetization field components must satisfy the Euler-Lagrange equations: $ \nabla^2 m^\alpha = 0, \; \alpha=\{x,y\}$,  in any region without defects. Dipolar fields are the leading contribution to the multipole solution of these equations. Now, if one has many defects randomly placed at positions $\vec{x}_i$ the fluctuations of the magnetization field is:
\begin{equation}
\label{eq:magfluc}
    \left(m^\alpha(\vec{r}) \right)^2 = \sum_{ij} \mu_{i,\beta}^\alpha M_{ij}^{\beta\gamma} \mu_{j,\gamma}^\alpha \; ,
\end{equation}
where
\begin{equation}
M^{\beta\gamma}_{ij} = \frac{(\vec{r}-\vec{x}_i)^\beta(\vec{r}-\vec{x}_j)^\gamma}{|\vec{r}-\vec{x}_i|^2|\vec{r}-\vec{x}_j|^2} \; ,
\end{equation}
and $\vec \mu^\alpha (\vec x)$ are the dipole moments, which we assume to be random vectors with $\mu^2 = (\mu^x)^2 = (\mu^y)^2$. By considering a finite density of defects $\rho(\vec{x})$, we permute sums with integrals in \eqref{eq:magfluc} and retaining only the diagonal elements, which are the dominating ones, we write
    \begin{equation}
    \left(m^\alpha(\vec{r}) \right)^2 \simeq \int d^2x~\rho(\vec{x})~\mu^\alpha_\beta(\vec{x}) M^{\beta\gamma}(\vec{x}) \mu^\alpha_\gamma(\vec{x}) \; .
\end{equation}
Solving this integral for a uniform density $\rho(\vec x) \sim p^2$ we find a logarithmic divergence:
\begin{equation}
    |\vec{m}|^2 \sim  2 \pi  \mu^2 p^2 \ln\left( \frac{\ell}{a} \right) \; ,
\end{equation}
where $a$ is the microscopic lattice spacing and $\ell \sim L$. This divergence signals the breaking of the original small fluctuation hypothesis, suggesting that magnetic order in $d=2$ may be unstable against any finite density of defects in the thermodynamic limit. For any arbitrarily small $p$, there is always a system size $L_{p}$ such that the cumulative effect of the defects at any point of the lattice is non-negligible. This lenghtscale sets the size of the magnetic domains and is exponentially large:
\begin{equation}
\label{eq:scale_classic}
    L_{p} \sim a\ e^{\frac{c}{p^2}} \; ,
\end{equation}
with $c \sim 1/(2\pi\mu^2) \sim 1$. For $p \sim 10^{-2}$, such domains are expected to be of size $L \sim 10^{40}$, making them larger than any experimental (or numerical) sample \cite{Jenkins2022}. 

It also has to be noticed that this computation neglects dipoles interactions \cite{Parker1988}, which may produce screening effects protecting the magnetization and possibly pushing $L_p$ to larger values or infinity. We have preliminary results regarding the inclusion of interactions, but there is no room for them in this paper.

\subsection{Derivation of the effective field theory of magnons}
The SW Hamiltonian matrices $A,B$ are correlated random matrices, whose entries depend on the specific realization of disorder $J_{ij}$ and its configuration of minimum. Computing its spectral properties necessarily require some exact diagonalization routine. However, we can, with some minimal assumptions, build a phenomenological field theory which describes the behavior of the SWs. Because of the global symmetry $A,B$ must be proportional to a discrete Laplacian term:
\begin{eqnarray}
H_{SW}&=\sum_{i}\rho^{(0)}_i(\nabla a^\dagger)_i(\nabla a)_i  \nonumber \\
&+\sum_{i}\frac{\theta^{(0)*}_i}{2}(\nabla a^\dagger)_i(\nabla a^\dagger)_i  \nonumber \\
&+\sum_{i}\frac{\theta^{(0)}_i}{2}(\nabla a)_i(\nabla a)_i \; ,
\end{eqnarray}
where $(\nabla a^\dagger)_i\cdot(\nabla a)_i=(a^\dagger_{i+e_x}-a_{i})(a^\dagger_{i+e_x}-a_{i})+(a^\dagger_{i+e_y}-a_{i})(a^\dagger_{i+e_y}-a_{i})$.
Passing to Fourier transform, in the limit in which $\rho^{(0)}_i,\theta_i$ are constant we have
\begin{equation}
H_{SW}=\sum_{\vec{k}}\epsilon_k\rho^{(0)}a^\dagger_{\vec{k}}a_{\vec{k}}+\frac{\theta^*}{2}a^\dagger_{\vec{k}}a^\dagger_{\vec{k}}+\frac{\theta}{2}a_{\vec{k}}a_{\vec{k}} \; ,
\end{equation}
with $\epsilon_k=2-\cos(k_x)-\cos(k_y)$ are the eigenvalues of the Laplacian on the square lattice (with $a=1$).

For $\theta\ll\rho$, which means $B\ll A$, which we verified numerically for small $p$, one can rotate to Bogoljubov bosons and get
\begin{equation}
H_{SW}=\sum_{k}\rho\;\epsilon_k b^\dagger_{\vec{k}}b_{\vec{k}}\; ,
\end{equation}
with $\rho=\sqrt{\rho^{(0)\; 2}-|\theta|^2}$. We can now pass to the continuum and re-introduce a smooth dependence of $\rho$ on $\vec x$, let $\phi$ be the continuum limit of the Bogoljubov boson $b$ and $\epsilon_k\sim \frac{k^2}{2}$. In this way we recover the Hamiltonian equivalent of \eqref{eq:ActionM}. 

\subsection{One-loop RG computation}
Let us here explicitly compute the integrals corresponding to the one-loop diagrams drawn in Fig.~\ref{fig:RGdiag}. We start by reminding the expressions of the bare propagators in momentum space, which are the building blocks of the computation. The bare propagator of the quenched disordered field $\sigma$ -- represented by a dashed line in the diagrams -- can be written as 
\begin{equation}
    \langle \sigma(\vec q, \Omega) \sigma(-\vec q', -\Omega) \rangle_0 = \frac{2\pi \delta(\Omega)}{q^2 + \mu^2} \; , 
\end{equation}
where $\delta(\Omega)$ explicitly expresses the fact that the disorder line is static, \emph{i.e.} it carries no frequency. The bare propagator of the field $\phi$ -- represented by a continuous line in the diagrams -- is the clean ferromagnetic propagator:
\begin{equation}
    G_0(q,\omega) \equiv \langle \phi(\vec q, \omega) \phi^*(-\vec q', -\omega) \rangle_0 = \frac{i/s}{\omega - \rho_s q^2} \; .
\end{equation}

For this one-loop computation, we use a UV cutoff regularization scheme. The self-energy can be then computed as:
\begin{widetext}
\begin{equation}
    \Sigma_1(p,\omega) = (-i g s)^2 \int^\Lambda \frac{d^2q}{(2\pi)^2} \frac{1}{q^2 + \mu^2} (\vec{p}\cdot(\vec{p}-\vec{q}))^2 G_0(\vec{p}-\vec{q},\omega) = \frac{i g^2 s}{\rho_s} I_{\Lambda,\mu}(\vec p,\omega) \; ,
\end{equation}
where we have defined the integral
\begin{equation}
\label{eq:int1loop}
    I_{\Lambda,\mu}(\vec p,\omega) = \int^\Lambda \frac{d^2q}{(2\pi)^2} \frac{1}{q^2 + \mu^2} \frac{(\vec{p}\cdot(\vec{p}-\vec{q}))^2}{(p-q)^2 - \omega/\rho_s } \; ,
\end{equation}
$\Lambda$ being the UV cutoff. By extracting the UV divergent part of this integral, we find the following expression of the one-loop self energy:
\begin{equation}
  \begin{split}
      \Sigma_1(p,\omega) & = \frac{i g^2 s}{4 \pi \rho_s} \ln \left(\frac{\Lambda}{\mu}\right) p^2 +\mathrm{finite} \; ,
  \end{split}  
\end{equation}
which is independent of $\omega$, meaning that the time derivative part of the action is not renormalized at one loop, \emph{i.e.} the field renormalization constant is $Z_\phi=1$. All 1PR (\emph{one particle reducible}) diagrams of this kind appearing at higher orders in the perturbative expansion can be resummed through Dyson equation:
\begin{equation}
    G_{\Lambda,\mu} = G_0 \sum_{n \geq 0} (\Sigma_1 G_0)^n = G_0 \frac{1}{1-\Sigma_1 G_0} = \frac{1}{G_0^{-1} - \Sigma_1} \; ,
\end{equation}
with $G_0^{-1} = -i s(\omega -\rho_s q^2)$. Eventually, the one-loop renormalized propagator has the following expression:
\begin{equation}
    G_{\Lambda,\mu}(q,\omega) = \frac{i/s}{\omega - \rho_s \left(1 - \frac{g^2}{4 \pi \rho_s^2} \ln (\Lambda/\mu) \right) q^2  } \; ,
\end{equation}
which yields the following correction to the stiffness:
\begin{equation}
    \rho_s \rightarrow \rho_s - \frac{g^2}{4 \pi \rho_s} \ln \frac{\Lambda}{\mu} \; .
\end{equation}

The vertex gets renormalized at order $g^3$, by the following integral:
\begin{equation}
\label{eq:vert}
        A_{\Lambda,\mu}(\vec{p},\vec{p}',\omega) =  (-i g s)^3 \int^\Lambda \frac{d^2 q}{(2\pi)^2} \frac{1}{q^2 + \mu^2} G_0(\vec{p}-\vec{q},\omega) G_0(\vec{p}' -\vec{q},\omega) \vec{p} \cdot (\vec{p} -\vec{q}) (\vec{p} -\vec{q}) \cdot  (\vec{p}' -\vec{q}) (\vec{p}' -\vec{q})\cdot \vec{p}' \; .
\end{equation}
Its UV divergent part can be easily extracted and its final expression is:
\begin{equation}
     A_{\Lambda,\mu}(\vec{p},\vec{p}',\omega) = - \frac{i g^3 s}{ 4 \pi \rho_s^2}  \ln \left(\frac{\Lambda}{\mu} \right) (\vec{p} \cdot \vec{p}')  + \mathrm{finite}  \; .
\end{equation}
Since the vertex function structure is $- i g s (\vec{p} \cdot \vec{p}')$, the disorder coupling receives the following correction:
\begin{equation}
    g \rightarrow g + \frac{g^3}{4 \pi \rho_s^2}\ln \frac{\Lambda}{\mu} \; .
\end{equation}

\end{widetext}

\bibliography{ref}

\end{document}